\documentclass[12pt]{article}
\usepackage{amsmath}
\usepackage{amssymb}
\usepackage{epsfig}
\usepackage{euscript}
\usepackage{fancybox}
\usepackage{color}
\def\mathswitchr#1{\relax\ifmmode{\mathrm{#1}}\else$\mathrm{#1}$\fi}

\newcommand {\pslash}{\hbox{$\not\hbox{\kern-2.3pt $p$}$}}

\usepackage{cite}

\usepackage{epic}
\def\rQCED{{\rm QCED}}

\def\alf1{ {\alpha\over\pi} }
\begin{document}
%\input{feynman} 
%=======================================================================
\begin{titlepage}
\begin{flushright}
%{\bf MPI-PhT-2002-08}\\
 {\bf BU-HEPP-03/10, UTHEP-03-1001 }\\
{\bf Oct., 2003}\\
\end{flushright}
%\vspace{0.05cm}
 
\begin{center}
{\Large QED$\otimes$QCD Threshold Corrections at the LHC$^{\dagger}$
}
\end{center}

\vspace{2mm}
\begin{center}
{\bf   C.Glosser$^{a,b}$, S. Jadach$^{c,d}$, B.F.L. Ward$^{a\ddagger}$ and S.A. Yost$^{a}$}\\
%%{\bf   B.F.L. Ward}\\
\vspace{2mm}
{\em $^a$Department of Physics,\\
  Baylor University, Waco, Texas 76798-7316, USA}\\
{\em $^b$Department of Physics and Astronomy,\\
  The University of Tennessee, Knoxville, Tennessee 37996-1200, USA}\\
{\em $^c$CERN, Theory Division, CH-1211 Geneva 23, Switzerland,}\\
{\em $^d$Institute of Nuclear Physics,
        ul. Radzikowskiego 152, Krak\'ow, Poland,}
%{\em $^c$Werner-Heisenberg-Institut, Max-Planck-Institut fuer Physik,
%Muenchen, Germany,}\\
%{\em $^a$Werner-Heisenberg-Institut, Max-Planck-Institut fuer Physik,
%Muenchen, Germany,}\\
%{\em $^d$Department of Physics and Astronomy,\\
%  The University of Tennessee, Knoxville, Tennessee 37996-1200, USA.}\\
%{\em $^c$SLAC, Stanford University, Stanford, California 94309, USA,}\\
%{\em Department of Physics and Astronomy,\\
%  The University of Tennessee, Knoxville, Tennessee 37996-1200, USA}\\
%{\em $^c$SLAC, Stanford University, Stanford, California 94309, USA,}\\
\end{center}

%\begin{center}
% {\bf S. Jadach}\\
%{DESY, Theory Division, D-22603 Hamburg, Germany}\\
%   {\em Institute of Nuclear Physics,
%        ul. Kawiory 26a, Krak\'ow, Poland}\\
%   {\em CERN, Theory Division, CH-1211 Geneva 23, Switzerland,}\\
% {\bf W.  P\l{a}czek}\\
%   {\em Institute of Computer Science,
%   Jagellonian University, ul. Nawojki 11, 30-072 Krak\'ow, Poland}\\
%{\em CERN, Theory Division, CH-1211 Geneva 23, Switzerland,}\\
% {\bf M. Skrzypek}\\
%   {\em Institute of Nuclear Physics,
%        ul. Kawiory 26a, Krak\'ow, Poland}\\
%{\em CERN, Theory Division, CH-1211 Geneva 23, Switzerland,}\\
% {\bf B.F.L. Ward}\\
%   {\em Department of Physics and Astronomy,\\
%   The University of Tennessee, Knoxville, Tennessee 37996-1200\\
%   SLAC, Stanford University, Stanford, California 94309}\\
%{\em CERN, Theory Division, CH-1211 Geneva 23, Switzerland,}\\
%{\bf Z. W\c as}\\
%   {\em Institute of Nuclear Physics,
%        ul. Kawiory 26a, Krak\'ow, Poland}\\
%   {\em CERN, Theory Division, CH-1211 Geneva 23, Switzerland}\\
%\end{center}

\vspace{5mm}
\begin{center}
{\bf   Abstract}
\end{center}
We use the theory of YFS resummation to compute the size of the
expected resummed soft radiative threshold effects in 
precision studies of heavy
particle production at the LHC, where accuracies of 1\% are
desired in some processes. We find that the soft QED threshold effects are
at the level of 0.3\% whereas the soft QCD threshold effects enter at the 
level of 20\% and hence both must be controlled to be on 
the conservative side to achieve such goals.
\vspace{10mm}
\vspace{10mm}
\renewcommand{\baselinestretch}{0.1}
\footnoterule
\noindent
{\footnotesize
\begin{itemize}
\item[${\dagger}$]
Work partly supported 
% the Polish Government
%grants KBN 2P30225206 and 2P03B17210, the Maria Sk\l{}odowska-Curie
%Joint Fund II PAA/DOE-97-316, and
by the US Department of Energy Contract  DE-FG05-91ER40627.
\item[${\ddagger}$]
On leave of absence from the Department of Physics and Astronomy, University of Tennessee, Knoxville, TN 37996-1200.
%, and by
%Polish Government grant 5P03B09320.
\end{itemize}
}
%\vspace{0.5cm}
%\begin{flushleft}
%{\bf UTHEP-00-0101}\\
%{\bf Jan, 2000}\\
%\end{flushleft}

\end{titlepage}

%=======================================================================
\def\Kmax{K_{\rm max}}\def\ieps{{i\epsilon}}\def\rQCD{{\rm QCD}}
\renewcommand{\theequation}{\arabic{equation}}
\font\fortssbx=cmssbx10 scaled \magstep2
\renewcommand\thepage{}
%\vfill\eject
\parskip.1truein\parindent=20pt\pagenumbering{arabic}\par
%\section{\bf Introduction}\label{intro}\par
%%%Start here 
The physics objectives of the LHC entail a precise knowledge of the
Standard Model processes that are either background to the 
would-be discovery processes, such as Higgs and susy processes, or 
important for the normalization of the observed signal and backgrounds
processes so that the cross sections of both types of processes
can be determined in a way which maximizes the physics output
of this pioneering high energy colliding beam device.
One of the many sources of uncertainty that must be controlled are
those then associated with higher order radiative corrections
to all aspects of the 15 TeV pp collisions. The dominant
source of these effects are the higher order QCD corrections and 
many authors~\cite{qcdlit} have worked and are still  
working on these corrections.
But at the level of precision required for some of the LHC objectives,
one must also check that the higher order EW corrections are
under the appropriate control. It is well-known that initial state
QED corrections can be large, particularly for resonate heavy particle production, and hence, already in Ref.~\cite{cern2000,spies} estimates were made of
the size of the QED corrections to the structure function evolution
for the LHC energies. An effect at the level of a few per mille was
found in most of the relevant range of the Bjorken variable $x$.
More recently, in Ref.~\cite{james1}, a similar result was found
with however a qualitatively different character --
whereas the results in Refs.~\cite{cern2000, spies} all show that
the QED correction decreases the structure functions in the 
relevant regime of $x$, those in Refs.~\cite{james1} actually 
show a QED correction
which changes sign in the relevant regime of $x$, but which is still
similar in magnitude, a few per mille in general in the relevant regime.
This could just be the difference in the bases used to present the 
respective results, as a recent more complete treatment of the
these same effects in Ref.~\cite{roth} confirms both of the 
behaviors found in Refs.~\cite{spies,james1}.
%While we feel this qualitative difference should eventually be understood,
In the current discussion, we do not address these effects further. 
Rather, we focus on the
potentially larger issue of the size of the non-universal QED corrections associated with the threshold behavior of the heavy particle production at the
LHC. To illustrate these effects, we treat as prototypical processes
the processes 
$pp\rightarrow V +n(\gamma)+m(g)+X\rightarrow \bar{\ell} \ell'
+n'(\gamma)+m(g)+X$, where 
$V=W^\pm,Z$,and $\ell = e,\mu,~\ell'=\nu_e,\nu_\mu ( e,\mu )$
respectively for $V=W^+ ( Z )$, and  
$\ell = \nu_e,\nu_\mu,~\ell'= e,\mu$ respectively for $V = W^-$.
These processes are potential luminosity processes for the
LHC~\cite{lhclum}\footnote{ We understand~\cite{fnallum} that FNAL 
experiments are also
considering these processes as a potential luminosity processes.}. 
What we want to do is to estimate the expected size of 
the threshold contributions from these
corrections in the realistic acceptance for the process in the 
LHC detectors. In this way, we show what role if any these 
threshold radiative effects may play
in precision LHC studies where the luminosity is expected to
be needed at the few percent precision level. We recall for reference
that the current precision on the FNAL luminosity 
is $\sim 6\%$~\cite{fnallum}. We stress that, for an overall precision
of 2-3\%, the theory budget in the error must be held to $\sim 1\%$
so that it does not figure too prominantly in the total LHC
luminosity budget.\par

Physically, the source of the effects which we calculate will be the
following. In the complete calculation, assuming we have a factorization
scale $\mu_F$ comparable to the renormalization scale $\mu_R$,
both of which are set near the heavy particle production scale
$M$, one has the big logs $L_s\equiv \ln M^2/\mu_s^2,~L_{th}\equiv 
L = \ln \hat s/M^2$, where here we assume with Refs.~\cite{cern2000,spies}
that the structure functions have been evolved from input data
at scale $\mu_s$ that have QED radiative effects in them.
$\hat s$ is the invariant mass squared of the hard parton production process.
\footnote{We stress that there are apparently data available~\cite{thorne1} 
from which the
QED corrections from quarks have been removed 
so that the attendant QED correction to the structure function evolution
would have a different boundary condition than that used
in Refs.~\cite{cern2000,spies}. We are not aware of any results
in the published literature where this type of calculation has been done.}
Hence, all of the familiar big logs of the type $\ln Q^2/m_f^2$
that one would expect from analogy with the big logs in
$e^+e^-$ annihilation at high energies are removed by the factorization
of the cross section's mass singularities as usual.
Here, Q is any hard scale and $m_f$ would be the respective
quark mass. What do remain are the residual big logs $L_s,L_{th}$,
where the former are presumably all summed up by the
structure function evolution. This leaves the latter, which we
now study using the YFS methods in Ref.~\cite{yfs,yfs1}.\par

We start from the basic formula
\begin{equation}
d\sigma_{exp}(pp\rightarrow V+X\rightarrow \bar\ell \ell'+X') =
\sum_{i,j}\int dx_idx_j F_i(x_i)F_j(x_j)d\hat\sigma_{exp}(x_ix_js),
\label{sigtot} 
\end{equation}
where the squared $pp$ cms energy is $s$ and the
structure functions have their usual meaning here so that
$\hat\sigma_{exp}(x_ix_js)$ is the reduced YFS exponentiated V production
cross section, $V=W,Z$. The YFS formula which we use in the 
latter cross section
was already developed in Ref.~\cite{top1}, where the 
dominant higher order corrections from the QCD corrections for
top production at FNAL were estimated. Here, we can extend that result by
including the dominant QED corrections in addition to the ISR
QCD corrections treated in Ref.~\cite{top1}.\par

We point out that the threshold resummation result
which we obtained in Ref.~\cite{top1} is agreement with the results
of Catani {\it et al.} in Ref.~\cite{catani} for the size
of the higher order threshold resummation effects in QCD for t\=t
production at the Tevatron. As we have explained in Ref.~\cite{dglapsyn},
this is not an accident. The 
Sterman-Catani-Trentadue (SCT)~\cite{ster,cattrent} 
resummation theory for QCD used in Ref.~\cite{catani} is fully 
consistent with the QCD extension of 
the YFS theory that we discuss in this paper. As we explain in
Ref.~\cite{dglapsyn}, one has to use the 
corresponding values of the $\bar\beta_n$ residuals in the 
extension of the YFS theory to QCD and the corresponding
higher order corrections to the infrared exponent $SUM_{ir}(QCD)$
in eqs.(15-18) in Ref.~\cite{dglapsyn} to get an exact correspondence between
the two resummation approaches. As we will see below, the result
which we obtain for the QCD resummed threshold effects in Z production
using our extension of the YFS methods to QCD will account for 
most of the actual QCD correction as given by the exact result 
through ${\cal O}(\alpha_s^2)$ in
Refs.~\cite{van1,van2}. This again shows the applicability of our
methods to this problem and, by our correspondence in Ref.~\cite{dglapsyn},
the applicability of the SCT method as well. 
We do not pursue the latter method further here.\par

Specifically, we recall the need to use 
Dokshitzer-Gribov-Lipatov-Altarelli-Parisi ( DGLAP ) 
synthesized~\cite{dglapsyn}
YFS formulas so that we avoid double counting of effects.
In practice, this amounts to applying the factorization theorem to the
YFS real and virtual infrared functions, removing those parts
of these functions which generate the big logs $L_s$.
The factorization scheme used should be that used in the
isolation of these big logs to the respective structure functions.
This means that the functions $B^{nls},~{\tilde B}^{nls},~{\tilde S}^{nls}$
defined in Ref.~\cite{dglapsyn} should be used in the YFS formulas
applied to the factorized reduced hard parton-parton
production cross section. Here, we stress that we are doing this for both
the QCD and the EW QED corrections.
This means that, in the basic YFS algebra illustrated
in Ref.~\cite{top1}, we need to make the replacements, using an obvious notation, 
\begin{eqnarray}
B^{nls}_{QCD} \rightarrow B^{nls}_{QCD}+B^{nls}_{QED}\equiv B^{nls}_{QCED},\cr
{\tilde B}^{nls}_{QCD}\rightarrow {\tilde B}^{nls}_{QCD}+{\tilde B}^{nls}_{QED}\equiv {\tilde B}^{nls}_{QCED}, \cr
{\tilde S}^{nls}_{QCD}\rightarrow {\tilde S}^{nls}_{QCD}+{\tilde S}^{nls}_{QED}\equiv {\tilde S}^{nls}_{QCED}
\end{eqnarray} 
in the YFS exponentiation algebra in Ref.~\cite{top1},
with the attendant replacement of the reduced parton cross sections
and structure functions to reflect the incoming $pp$ state as it is
compared to the incoming $p\bar p$ state in Ref.~\cite{top1}
and to reflect the V production and decay here versus the
$t\bar t$ production there.\par

In this way, we start from the basic result
\begin{eqnarray}
d\hat\sigma_{\rm exp} = e^{\rm SUM_{IR}(QCED)}
   \sum_{{n,m}=0}^\infty\int\prod_{j_1=1}^n\frac{d^3k_{j_1}}{k_{j_1}} \cr
\prod_{j_2=1}^m\frac{d^3{k'}_{j_2}}{{k'}_{j_2}}
\int\frac{d^4y}{(2\pi)^4}e^{iy\cdot(p_1+q_1-p_2-q_2-\sum k_{j_1}-\sum {k'}_{j_2})+
D_\rQCED} \cr
\tilde{\bar\beta}_{n,m}(k_1,\ldots,k_n;k'_1,\ldots,k'_m)\frac{d^3p_2}{p_2^{\,0}}\frac{d^3q_2}{q_2^{\,0}},
%\end{split}
\label{subp15}
\end{eqnarray}
where  the new YFS residuals $\tilde{\bar\beta}_{n,m}(k_1,\ldots,k_n;k'_1,\ldots,k'_m)$, with $n$ hard gluons and $m$ hard photons,
represent the successive application
of the YFS expansion first for QCD as in Ref.~\cite{top1,app1,ichep02}
followed by that for QED as some of us have used in many applications
as in Refs.~\cite{yfs1}.
The infrared functions are now given by
\begin{eqnarray}
{\rm SUM_{IR}(QCED)}=2\alpha_s\Re B^{nls}_{QCED}+2\alpha_s{\tilde B}^{nls}_{QCED}\cr
D_\rQCED=\int \frac{dk}{k^0}\left(e^{-iky}-\theta(K_{max}-k^0)\right){\tilde S}^{nls}_{QCED}
\label{irfns}
\end{eqnarray}
where the dummy parameter $K_{max}$ is just that -- nothing depends on it.
We have taken the liberty, not to be viewed as permanent, to set
the dummies in QCD and QED to be the same value. The infrared algebra
realized here for QCD and QED we will sometimes denote as 
Quantum ChromoElectroDynamics (QCED).\par

The result (\ref{subp15}) may be realized in the context
of (\ref{sigtot}) using Monte Carlo methods as illustrated in
Refs.~\cite{ichep02}. We shall present those methods in detail elsewhere.
Here, to illustrate the size of the QED effects in the presence
of the QCD resummation as well, we use a semi-analytical evaluation
of (\ref{sigtot}).\par

Specifically, in order isolate explicitly ISR effects, we take the
case of single Z production and refer the case of single W
production to Ref.~\cite{elswh}. We also call attention to the complete
EW ${\cal O}(\alpha)$ corrections to single heavy gauge boson
production in hadron colliders in Refs.~\cite{baur0,baur,hol,ditt,russ}.
We work in the approximation that, when we retain the dominant infrared effects
at the $\tilde{\bar\beta}_{0,0}^{(0,0)}$-level,
these two sets of corrections commute with one another:
physically, gluons are EW singlets and photons are color singlets.
Thus, the only issue is whether particles such as quarks
with both color and EW charge have significant higher order corrections in 
which the two sets of corrections do not commute. Here, we appeal to the
Heisenberg uncertainty principle and note that the average 
soft photon energy fraction in the
YFS formalism is $x_{avg}(QED)\cong \gamma(QED)/(1+\gamma(QED))$
and the average YFS soft gluon energy fraction is correspondingly 
 $x_{avg}(QCD)\cong \gamma(QCD)/(1+\gamma(QCD))$, where the YFS
radiation probability strengths are given 
by $\gamma(A)=\frac{2\alpha_{A}{\cal C}_A}{\pi}(L_s
-1)$, $A=QED,QCD$. For quarks, ${\cal C}_A=Q_f^2, C_F$, respectively, for 
$A=QED,QCD$, where $Q_f$ is the quark electric charge and $C_F$ is
the quadratic Casimir invariant for the quark color representation.
Here, we variously use $\alpha_{QED}\equiv \alpha$,
~$\alpha_{QCD}\equiv \alpha_s$.
As the ratio of the two fine structure constants
is $\sim 10$ in the LHC/Tevatron environment in which we work, we expect that
the typical time for the dominant QCD corrections we exponentiate to
occur is more than an order of magnitude shorter than the corresponding time
for the analogous QED corrections which we exponentiate. Thus,
we do not expect that significant effects in our leading exponentiated analysis
are missing due to sets of QCD and EW QED corrections that do not 
commute. We then compute the 
ratio $r_{exp}=\sigma_{exp}/\sigma_{Born}$ with and without the
QED contributions, where, for definiteness, we note that the 
partonic Born cross section
we use here is well-known and we generate it by the 
standard methods from eq.(43) in Ref.~\cite{born1},
with the understanding that we retain only the Z contribution 
and that we substitute the respective incoming quark chiral couplings
for the incoming $e-$ chiral couplings accordingly.
(We stress that we {\em do not} use the narrow resonance approximation here.)
We get the results
\begin{equation}
r_{exp}=
\begin{cases}
1.1901&, \text{QCED}\equiv \text{QCD+QED,~~LHC}\\
1.1872&, \text{QCD,~~LHC}\\
1.1911&, \text{QCED}\equiv \text{QCD+QED,~~Tevatron}\\
1.1879&, \text{QCD,~~Tevatron}\\
\end{cases}
\label{res1}
\end{equation}
We see that indeed the QED threshold correction is at the level of 0.3\%
at both the Tevatron and at the LHC and is in principle significant 
for studies at the percent level
such as those that are envisioned for the precision luminosity
determination for the LHC~\cite{lhclum}\footnote{ We checked that the variation of the renormalization scale, which we set equal to the factorization
scale, above and below its nominal value $M_Z$ by a few widths of the Z does
not change our result that the QED ISR threshold correction enters at the .3\% level}. We also
corroborate here the well-known~\cite{van1,van2} result that the
QCD threshold correction is needed through ${\cal O}(\alpha_s^2)$
for such studies. The results in (\ref{res1}) are entirely consistent with the
exact ${\cal O}(\alpha_s^2)$ results in Ref.~\cite{van1,van2},
where comparison with the pure QCD results in eq.(\ref{res1})
shows that our threshold effects reproduce most of the complete
exact results at both the LHC and the Tevatron. 
The size of our exponentiated QED effects in (\ref{res1}) is fully 
consistent with the exact ${\cal O}(\alpha)$ results in 
Refs.~\cite{baur0,baur,hol}, where the ISR QED 
correction was found to be 0.4\%.
In arriving at our results in (\ref{res1}), we have 
used the parton distributions
from Ref.~\cite{mrst1}.\par

Our result for the size of the QED threshold effect is similar to the
size found in Refs.~\cite{cern2000,spies,james1,roth} for the 
size of the QED effects
on the structure function evolution itself. From these two results,
we can conclude that the ( higher order ) QED correction to heavy
particle production at the LHC will be at the several per mille level
and it can not really be ignored for percent level precision studies.\par
 
In conclusion, we have introduced a new theoretical construct in this
paper, QCED, the regime of the large IR effects in the 
large momentum transfer regime, where we expect the QED and QCD exponentiation
algebras to commute to a large extent and to the extent that they do
we identify the resulting theory as Quantum ChromoElectroDynamics.
Corrections to it are then to be treated perturbatively
in the respective combined YFS residuals $\tilde{\bar\beta}_{n,m}^{QCED}$.
We have used this paradigm to estimate the size of the QED
threshold effects in single gauge boson production at
the LHC type energies. We find that such effects are
large enough that they must be taken into account for precision
studies, such as those envisioned for precision luminosity 
determinations.\par
%%%START HERE

\section*{Acknowledgements}

We thank Profs. S. Bethke and L. Stodolsky for the support and kind
hospitality of the MPI, Munich, while a part of this work was
completed. We thank Prof. C. Prescott for
the kind hospitality of SLAC Group A while this
work was in progress.

\end{document}